\pdfoutput=1
\documentclass[prb,superscriptaddress,papersize=a4paper,twocolumn]{revtex4-1}
\usepackage{amsmath,amsfonts,amssymb,amsthm}
\usepackage{mathtext,mathtools,mathrsfs}
\usepackage{esint,bm,esvect,dsfont}
\usepackage{graphicx,epsfig,psfrag,xcolor,subcaption}
\usepackage{array,tabularx,tabulary,booktabs,multirow}
\graphicspath{{figures/}}
\usepackage{hyperref}
\hypersetup{unicode=true,
	pdftitle={Landau levels and magnetoconductance in Weyl semimetal p-n junction},
	pdfauthor={David Saykin}, pdfsubject={Scientific article},
	pdfcreator={David Saykin}, pdfproducer={David Saykin},
	pdfkeywords={Weyl semimetals} {Landau levels} {junction},
	colorlinks=true, linkcolor=blue, citecolor=green, urlcolor=blue,
	pdfstartview=FitH
}
\newcommand{\pn}{\ensuremath{p\text{\ensuremath{-}}n} }
\newcommand{\bigo}[1]{\ensuremath{\mathop{}\mathopen{}\mathcal{O}\mathopen{}\displaystyle\left(#1\right)}}
\newcommand{\const}{\ensuremath{\mathrm{const}}}
\newcommand{\figref}[1]{FIG.~\ref{#1}}
\newcommand{\tabref}[1]{TABLE~\ref{#1}}
\renewcommand{\H}{\ensuremath{\mathcal{H}}}

\begin{document}

\title{Landau levels with magnetic tunnelling in Weyl semimetal and\\ magnetoconductance of ballistic \pn junction}	
	\author{D.~R.~Saykin}
		\affiliation{Moscow Institute of Physics and Technology, Russia}
		\affiliation{Landau Institute for Theoretical Physics, Russia}
	\author{K.~S.~Tikhonov}
		\affiliation{Landau Institute for Theoretical Physics, Russia}
		\affiliation{National University of Science and Technology MISIS, Moscow, 119049 Russia}
	\author{Ya.~I.~Rodionov}
         \affiliation{Institute for Theoretical and Applied Electrodynamics, Moscow, 125412 Russia}
	\affiliation{National University of Science and Technology MISIS, Moscow, 119049 Russia}
\date{\today}

\begin{abstract}
We study Landau levels (LLs) of Weyl semimetal (WSM) with two adjacent Weyl nodes. We consider different orientations $\eta=\angle(\mathbf{B},\mathbf{k}_0)$ of magnetic field \(\mathbf{B}\) with respect to $\mathbf{k}_0$, the vector of Weyl nodes splitting. Magnetic field facilitates the tunneling between the nodes
giving rise to a gap in the transverse energy of the zeroth LL. We show how the spectrum is rearranged at different $\eta$ and how this manifests itself in the change of behavior of differential magnetoconductance $dG(B)/dB$  of a ballistic \pn junction. Unlike the single-cone model where Klein tunneling reveals itself in positive $dG(B)/dB$, in the two-cone case $G(B)$ is non-monotonic with maximum at $B_c \propto \Phi_0k_0^2/\ln(k_0l_E)$ for large $k_0l_E$, where $l_E=\sqrt{\hbar v/|e|E}$ with $E$ for built in electric field and $\Phi_0$ for magnetic flux quantum.
\end{abstract}
\pacs{NaN}

\maketitle

\section{Introduction}
Since the discovery of time-reversal invariant topological insulators (see Ref.~\onlinecite{KaneMele} and references therein) topological properties of the electronic band structure of crystalline materials have been enjoying a lot of attention. More recently it was demonstrated that topological properties are shared by accidental point band touchings so that they may become nontrivial and robust under broken either time reversal or inversion symmetry. The band structure near these points can be described by a massless two-component Dirac or Weyl Hamiltonian. After Ref.~\onlinecite{Wan2011} indicated the possibility of a WSM state for pyrochlore iridates, the quest for model Hamiltonians and material candidates ensued\cite{Burkov2011weyl,Burkov2011Topological,Halasz2012,Vafek2014}.

Experimentally, the WSM state was first discovered in TaAs\cite{Lv2015TaAsFermiArcs,Lv2015TaAs}  and TaP\cite{Xu2016TaP}, materials with the broken inversion symmetry. First principle calculations~\cite{Weng2015} revealed that in both materials all Weyl nodes form a set of closely positioned pairs of opposite chirality in momentum space, this prediction was confirmed by experimental observation. Recently, the active experimental research\cite{Jeon2014,Xiong2015,Huang2015,Li2016,Xu2015,Xu2015NbAs,Xiong2015ChiralAnomaly,Arnold2016,Murakawa1490,Wang2016} shifted from the initial band-structure study to the surface and transport phenomena: a significant amount of attention was devoted  to magnetotransport, which was addressed theoretically\cite{Spivak2013,Klier2015,Andreev2016,Spivak2016} and experimentally\cite{He2014,Zhao2015,Novak2015,Du2016}.

One of the most impressive manifestations of gapless band structure is Klein tunneling, which reveals itself in transport through a \pn junction. Such junctions have been investigated theoretically for graphene\cite{Cheivanov2006,Shytov2007,Zhang2008}, carbon nanotubes\cite{Andreev2007,Chen2010} and surface of topological insulators\cite{Wang2012}.
The recent study Ref.~\onlinecite{Andreev2016} was devoted to magnetoconductance of a \pn junction realized in WSM.
The authors showed that in the case of a longitudinally aligned external magnetic field: \(\mathbf{B}\parallel\mathbf{E}\) where \(\mathbf{E}\) is a junction's built-in electric field, the differential magnetoconductance $dG(B)/dB$ is positive. This situation is different from the ordinary semiconductor \pn junction\cite{Keldysh1958}, where the real tunneling between valence and conductance bands is involved and magnetoconductance is negative\cite{Aronov,Haering1962}. The authors of Ref.~\onlinecite{Andreev2016} argued that the positivity of  magnetoconductance is attributed to the existence of a zero mode in electron LLs with degeneracy $\propto B$ and field-independent reflectionless Klein tunneling.

The treatment of Ref.~\onlinecite{Andreev2016} was done in the approximation of well-separated (in the momentum space) Weyl nodes.
This is a standard approximation which holds well in many situations. Usually, the influence of pairwise~\cite{NielsenNinomya1983} structure of WSM nodes on transport phenomena is accounted for by simple multiplication of a single point contribution by the number of cones in the spectrum. This approach however, breaks down in strong magnetic fields. When cyclotron radius of a particle $R_c \sim \hbar c k_0/|e|B$ with the momentum $k_0$ becomes of the order of its coordinate uncertainty $k_0^{-1}$ or, in other words, characteristic length of motion $l_B = \sqrt{\hbar c/|e|B}$ becomes of order $k_0^{-1}$, internode coupling must be taken into account. For example, in TaAs~\cite{Lv2015TaAs} the momentum distance between the Weyl nodes in a pair is $2k_0=0.0183\ \AA^{-1}$ and it happens at fields of order $B \simeq \Phi_0k_0^2 \simeq 17$ T. Indeed, such field-induced tunneling between two nodes in a pair has already been observed experimentally \cite{Zhang2017}. Therefore, it is not sufficient to consider the problem in a single Weyl cone approximation in such relatively high fields.
On the one hand, taking into account of the full spectrum of WSM which has 12 pairs of Weyl nodes (like in TaAs) is an intractable problem. On the other hand, the distance between pairs of Weyl nodes in momentum space fortunately happens to be much larger than the distance between the nodes in a pair~\cite{Lv2015TaAsFermiArcs}. Therefore, the correct treatment is to consider the nodes pairwise.

The problem of field-induced inter-node tunneling was addressed in a semiclassical approximation~\cite{OBrien2016}, where the behavior of the carrier density of states for high-lying LLs was explored. Recently, the same problem was studied numerically\cite{Zhang2017,Lee2017}  for all LLs for the perpendicular orientation of field with respect to nodes splitting $\mathbf{k}_0$. It was discovered that the magnetic-induced tunneling
opens a gap in a LL zero mode. The numerical analysis shows that the gap is  non-perturbative in external magnetic field $B$.

\begin{figure*}[t]
    \centering
    \begin{subfigure}{0.32\textwidth}			
    	\includegraphics[width=\textwidth]{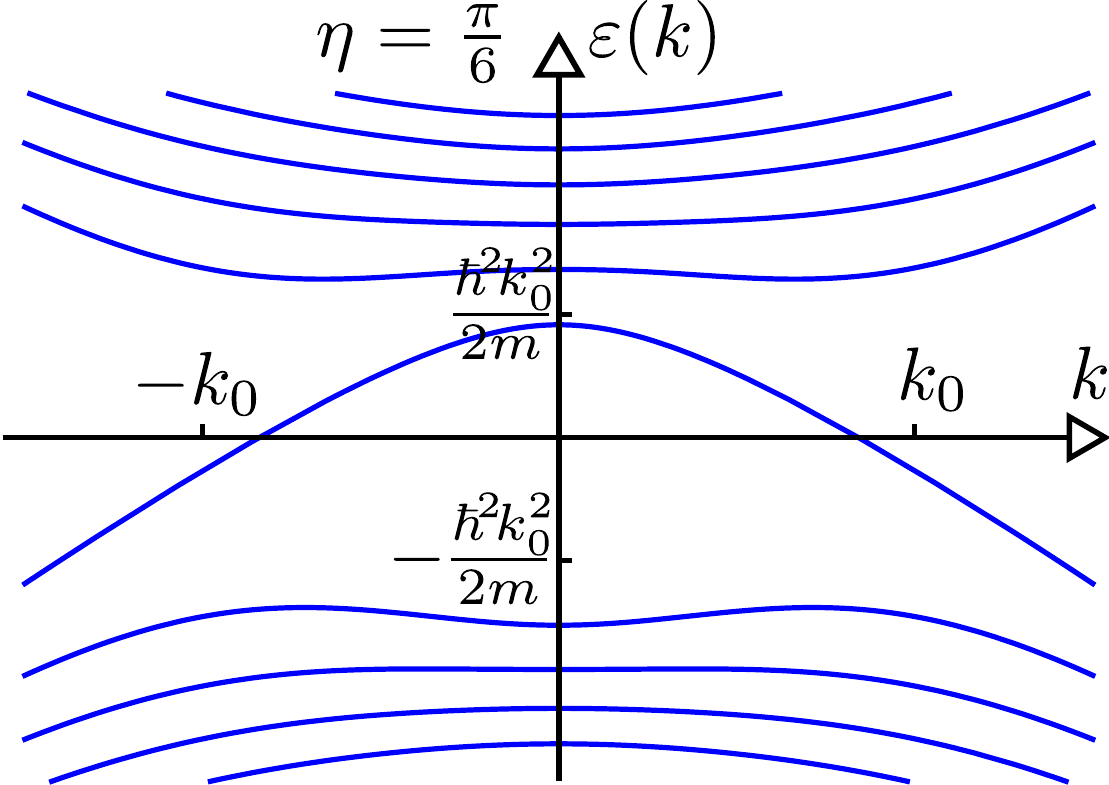}
    \end{subfigure}
    \enskip
    \begin{subfigure}{0.32\textwidth}			
    	\includegraphics[width=\textwidth]{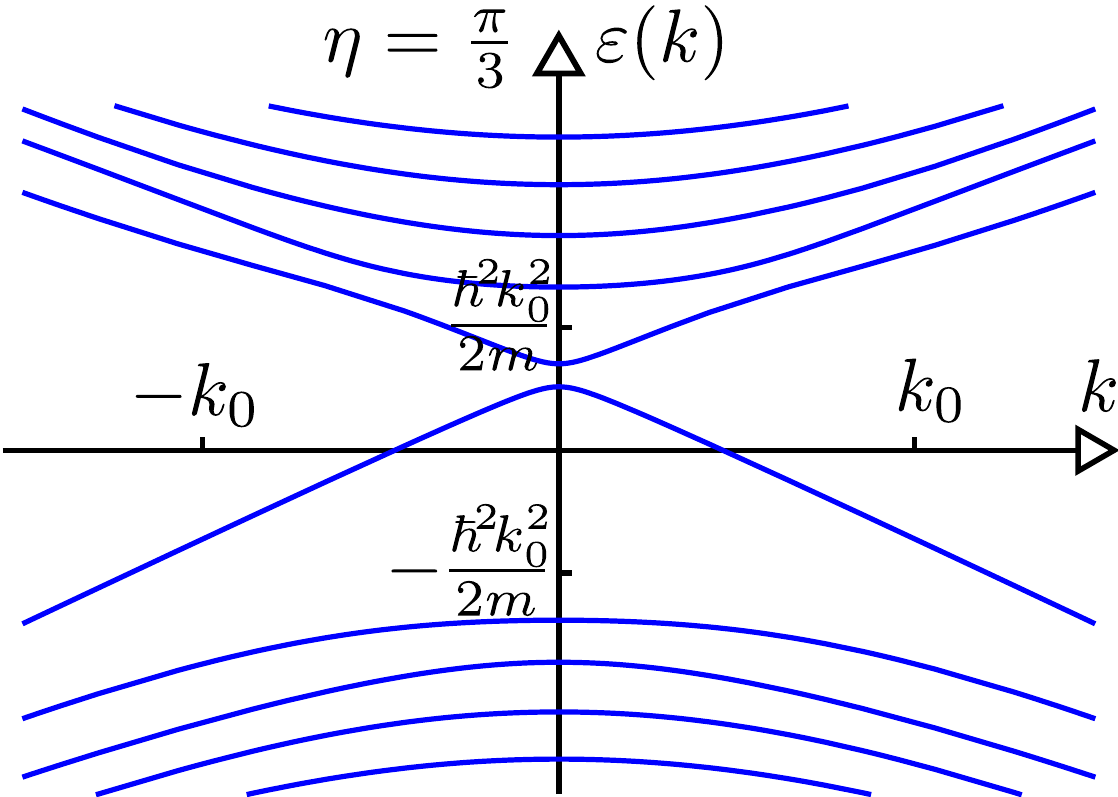}
    \end{subfigure}
    \enskip
    \begin{subfigure}{0.32\textwidth}		
    	\includegraphics[width=\textwidth]{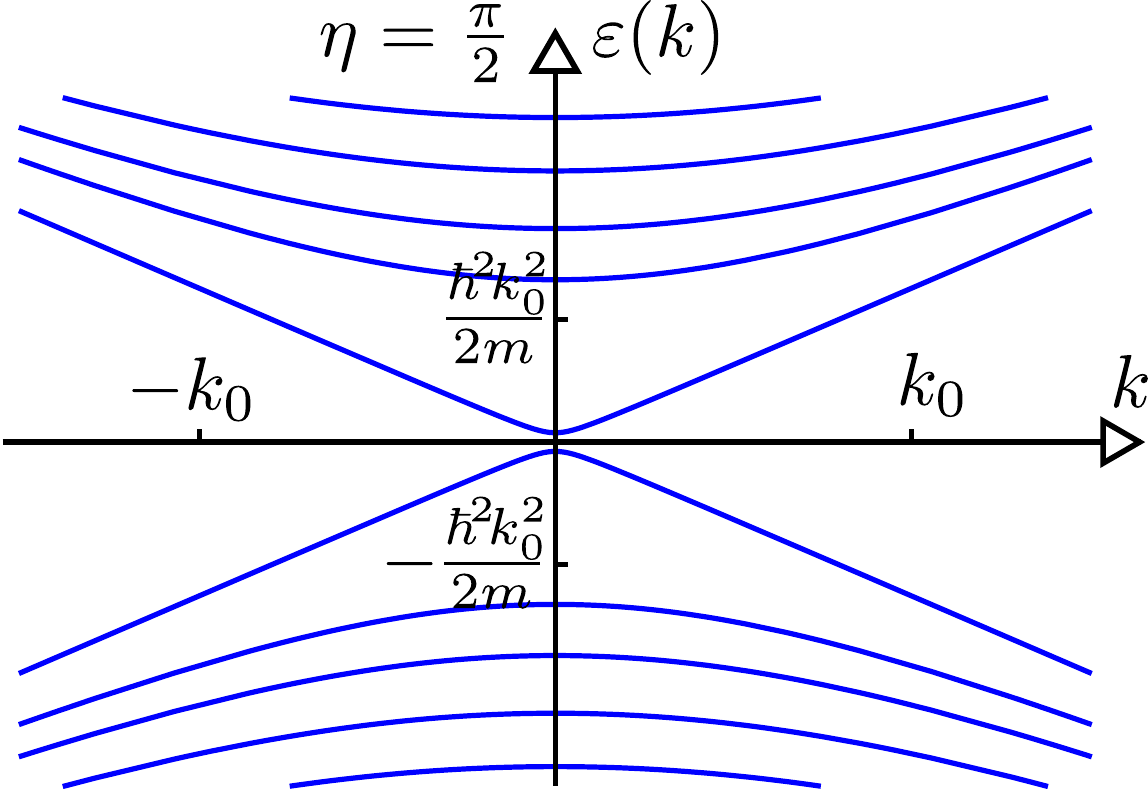}
    \end{subfigure}
    \caption{Energy versus momentum along the magnetic field $\mathbf{k}\parallel\mathbf{B}$ dependence (computed numerically) with parameters $\zeta = 1.6$ and $k_0l_B = 1.4$ for different angles $\eta = \pi/6$, $\pi/3$, $\pi/2$. Material parameters are taken for TaAs $\hbar^2k_0^2/2m \simeq 12$ meV and magnetic field is set to $B \simeq 34$ T to make the gap visible at $\eta=\pi/2$.}
    \label{fig:landau_levels}
\end{figure*}

We present here analytical theory of the spectrum of the LLs and its dependence on angle $\eta$ between magnetic field $\mathbf{B}$ and $\mathbf{k}_0$.
In particular, we show that for $\eta=\pi/2$ the problem of LLs  is reduced to the supersymmetric quantum mechanics of a particle in quadratic superpotential. Let us suppose the spectrum consists of two cones separated by the distance $2k_0$. Generic low--energy Hamiltonian for such system was derived in Ref.~\onlinecite{Okugawa2014}:
\begin{equation}
	\H = \Delta + \frac{\hbar^{2}}{2m} \left(\hat{k}_{x}^{2} - k_{0}^{2}\right)\sigma_{x}
	+ \hbar v \left( \hat{k}_{y}\sigma_{y} + \hat{k}_{z}\sigma_{z} \right),
    \label{eq:weyl_pair}
\end{equation}
where $\Delta$ is Weyl node energy measured from the chemical potential and $v$ stands for Fermi velocity. In what follows, we drop the energy offset $\Delta$ from most of the equations (it matters only in estimation of heterojunction's built--in potential, see in Ref.~\onlinecite{Andreev2016}). In the framework of a model described by Hamiltonian Eq. (\ref{eq:weyl_pair}), our results are as follows.
The lowest LL energy is indeed exponentially close to zero  and non-perturbative in magnetic field.
\begin{figure}[t!]
    \centering
   \includegraphics[width=0.9\columnwidth]{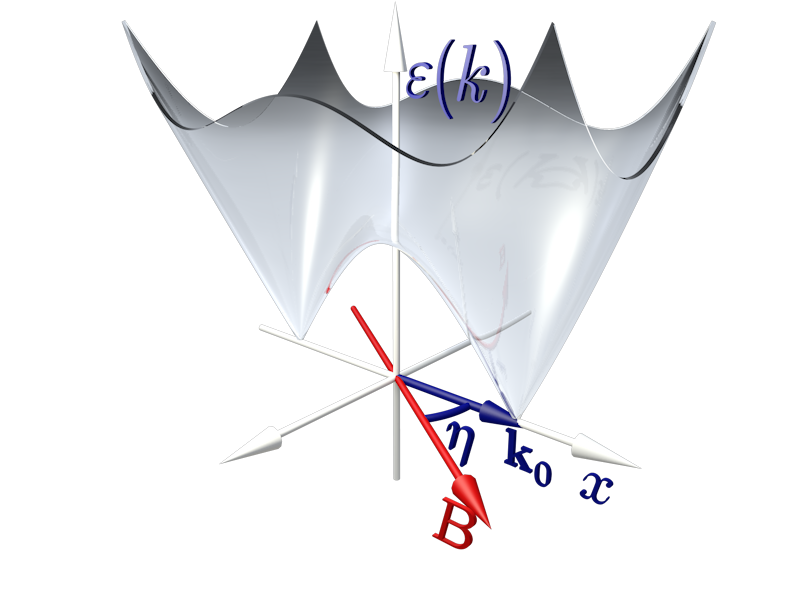}
    \caption{The orientation of Weyl spectrum and magnetic field. }
    \label{orient}
\end{figure}
\begin{figure}[t]
	\centering
	\includegraphics[width=.85\columnwidth]{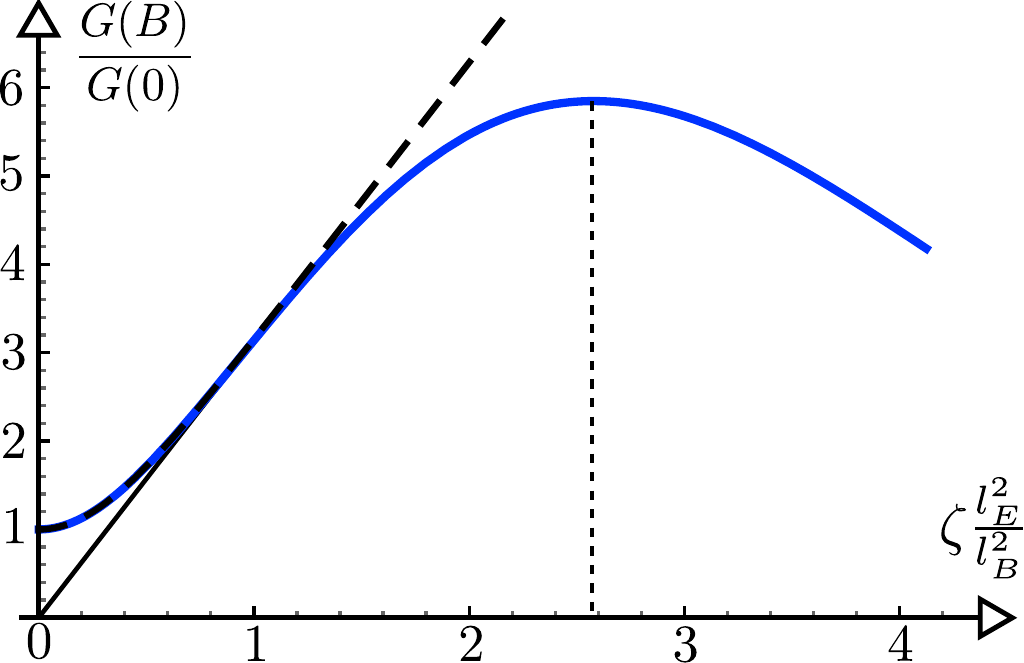}
	\caption{Magnetoconductance $\frac{G(B)}{G(0)}$ vs Magnetic field $\zeta \frac{vB}{cE}$ with parameter $\zeta k_0 l_E = 2$. Dashed line represents result of Ref.~\onlinecite{Andreev2016} and black line depicts first term of the sum \eqref{eq:sum}.}
	\label{fig:magnetoconductance}
\end{figure}
The exact formula for transversal energy reads
\begin{gather}\label{ground}
	\varepsilon_0 = \frac{(\hbar k_0)^2}{m}\sqrt{\frac{B}{\pi B_0}} \exp\left(-\frac{2B_0}{3B}\right),\quad B_0 = \zeta\frac{\Phi_0 k_0^2}{\pi},
\end{gather}
where $\zeta \equiv \frac{v_x}{v_y} = \frac{\hbar k_0}{mv}$ is anisotropy parameter. Numerically computed dependence of LLs on angle $\eta$ between magnetic field $\mathbf{B}$ and $\mathbf{k}_0$ is shown on the \figref{fig:landau_levels}.

We have also studied analytically the  magnetoconductance of WSM-based \pn junction for $\mathbf{E}\parallel\mathbf{B}\perp\mathbf{k_0}$ and found that the magnetoconductance becomes a non-monotonic function of $B$ (see. \figref{fig:magnetoconductance}). We found the field corresponding to the maximum of $G(B)$ to be equal to
\begin{equation}
	\label{eq:crit}
  B_c \sim\frac{2}{3} \frac{B_0}{\ln(\zeta k_0 l_E)},
	\quad	\zeta k_0 l_E \gg 1.
\end{equation}
where $l_E=\sqrt{\hbar v/|e|E}$ is the electric field length.

The paper is organized as follows: section II explores the structure of LLs in the two--conical system of Eq. (\ref{eq:weyl_pair}), section III is dedicated to the magnetoconductance of a ballistic \pn junction realized in such system and the conclusions are drawn in section IV.

\section{Landau levels}
We begin our analysis with search for energy dispersion law in the presence of the magnetic field starting from Hamiltonian~(\ref{eq:weyl_pair}). In the presence of magnetic field, Hamiltonian~(\ref{eq:weyl_pair}) can't be split into two independent parts of opposite chirality (only effectively, at $k_0\rightarrow\infty$) so the nodes should be treated simultaneously. We orient the coordinates so that $x$ axis points in the direction of Weyl node separation $\mathbf{k}_0$.
The magnetic field $\mathbf{B}=B(\cos\eta,0,\sin\eta)$ is inclined at the angle $\eta$ with respect to $x$ axis (see. \figref{orient}),
so that the field is described by potential $\mathbf{A}=B(-y\sin\eta,0,y\cos\eta)$.
At first, we solve an eigenvalue problem in two limiting cases $\eta=0$, $\frac{\pi}{2}$ analytically and then provide numerical solutions for arbitrary angles.\\
\emph{Field parallel to nodes splitting}.
We start from the case $\eta=\angle(\mathbf{B},\mathbf{k}_0)=0$. After the shift of the variable $y\mapsto y-k_{z}l_B^2$,  and unitary rotation $\psi\mapsto\frac{1}{\sqrt{2}}(1+i\sigma_{y})\psi$ the Hamiltonian transforms to
\begin{equation}
	\H = \hbar v\begin{pmatrix}
			-\frac{\hbar}{2mv}(k_{x}^{2}-k_{0}^{2}) & yl_B^{-2}-\partial_{y} \\
			yl_B^{-2}+\partial_{y} & \frac{\hbar}{2mv}(k_{x}^{2}-k_{0}^{2})
		\end{pmatrix}.
\end{equation}
Eigenfunctions are expressed through Hermite functions
$\psi_{n}^{\text{osc}}(y) = (2^nn!\sqrt{\pi}l_B)^{-1/2}e^{-y^2/2}H_n(y)$ as
\begin{equation}\label{eq:ll_parallel_psi}
	\psi_{n \neq 0} =
		\begin{pmatrix}
			c_{n}^{1}\psi_{|n|}^{\text{osc}}(yl_B^{-1}) \\
			c_{n}^{2}\psi_{|n|-1}^{\text{osc}}(yl_B^{-1})
	\end{pmatrix},
	\quad
    \psi_{0} =
		\begin{pmatrix}\psi_{0}^{\text{osc}}(yl_B^{-1})\\ 0
		\end{pmatrix},
\end{equation}
where $n\in\mathbb{Z}$ and coefficients $c_{n}^{1,2}$ are determined as normalized solutions of eigenproblem
\begin{equation}
	\begin{pmatrix}
		-\frac{\hbar^2}{2m}(k_{x}^{2}-k_{0}^{2}) & \hbar vl_B^{-1}\sqrt{2|n|}\\
		\hbar vl_B^{-1}\sqrt{2|n|} & \frac{\hbar^2}{2m}(k_{x}^{2}-k_{0}^{2})
	\end{pmatrix}
	\begin{pmatrix}
		c_{n}^{1}\\
		c_{n}^{2}
	\end{pmatrix}
	=
	\varepsilon_{n}
	\begin{pmatrix}
		c_{n}^{1}\\
		c_{n}^{2}
	\end{pmatrix}.
\end{equation}

\begin{figure}[!t]
    \centering
    \includegraphics[width=.8\linewidth]{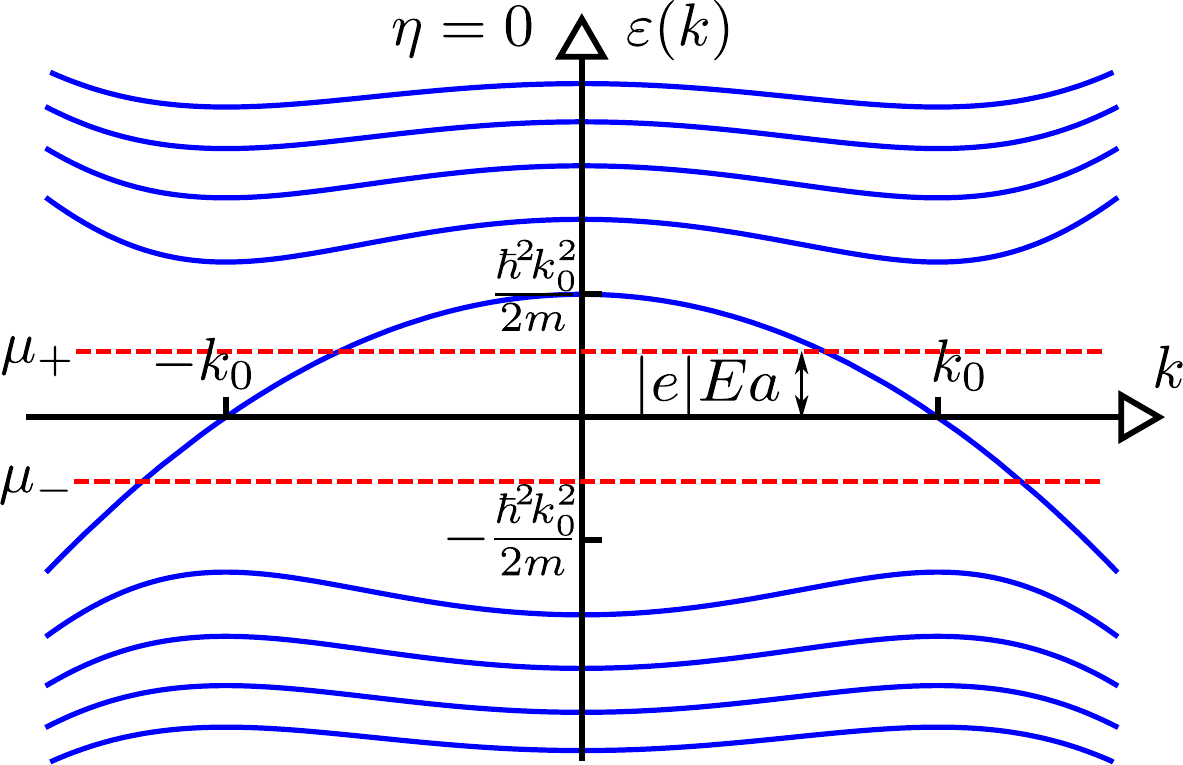}
    \caption{Landau levels for $\eta=\angle(\mathbf{B},\mathbf{k}_0)=0$ plotted from Eq. \eqref{eq:dispersion00}. Magnetic field is supposed to satisfy $l_Ba < l_E^2$ and electrochemical potentials from different sides of the junction are noted as $\mu_\pm=\mu(\pm\infty)$.}
    \label{fig:landau_levels_00}
\end{figure}

As a result, we find
\begin{equation}
\begin{aligned}
	\varepsilon_{n\neq0}(k_{x}) &= \hbar v\cdot \mathrm{sgn}(n)\sqrt{\frac{2|n|}{l_B^2}+\left(\frac{\hbar(k_{x}^{2}-k_{0}^{2})}{2mv}\right)^{2}},\\
	\varepsilon_{0}(k_{x}) &= -\frac{\hbar^2}{2m}\left(k_{x}^{2}-k_{0}^{2}\right),\qquad n\in\mathbb{Z}.
    \label{eq:dispersion00}
\end{aligned}
\end{equation}
Here we observe that spectrum~\eqref{eq:dispersion00} possesses electron-hole symmetry  for all non zero modes $n\rightarrow -n$.
However, the symmetry is violated for zero mode $n=0$ where only the hole state with negative energy exists. To make the whole picture  more intelligible we present the plot of LLs on \figref{fig:landau_levels_00}. The mentioned asymmetry, as the reader might have already guessed, is a remnant of the chiral anomaly. It is important to note that the zeroth LL is independent of magnetic field. It leads to
linear in $B$ magnetoconductance $G(B)$ for large $B$ in such orientation, see Section III. A similar scheme of LLs was obtained in paper Ref.~\onlinecite{Lu2015} for $\eta=0$ and almost identical Hamiltonian. Here we go further and retrieve LLs for any orientation of magnetic field with respect to $k_0$ vector.

\emph{Field perpendicular to nodes splitting.}
For the case $\eta=\angle(\mathbf{B},\mathbf{k}_0)=\frac{\pi}{2}$ after the shift $y\mapsto y+k_{x}l_B^2$ the Hamiltonian becomes
\begin{gather}
\label{HQ}
  \begin{split}
	\H &=\hbar v\begin{pmatrix}k_{z} & Q\\
	Q^\dag & -k_{z}
	\end{pmatrix},\\
 Q&=W-i\widehat{p}_y,\ \ W=\frac{\hbar}{2mv}\Big(\frac{y^{2}}{l_B^4}-k_{0}^{2}\Big).
  \end{split}
\end{gather}
To find the eigenvectors one may factorize $y$--dependent part of $\psi$--function as the solution of
\begin{gather}
  \begin{matrix}
 Q\chi^2_{n} & = -l_B^{-1}\epsilon_{n}\chi^1_{n}\\
 Q^\dag\chi^1_{n} & = -l_B^{-1}\epsilon_{n}\chi^2_{n}\\
  \end{matrix},\ \  \psi_{n} =
	\begin{pmatrix}
		c_{n}^{1}\chi^1_{|n|}(\frac{y}{l_B})\\
		c_{n}^{2}\chi^2_{|n|}(-\frac{y}{l_B})
	\end{pmatrix}.
      \label{eq:transverse_chi}
\end{gather}

When such vector $\chi^{1,2}_n(y)$ is found, eigenfunctions can be determined from the following linear equation
\begin{gather}
	\begin{pmatrix}
		k_{z}-\varepsilon_{n}(k_z) & -\epsilon_{|n|}\\
		-\epsilon_{|n|} & -k_{z}-\varepsilon_{n}(k_z)
	\end{pmatrix}
	\begin{pmatrix}
		c_{n}^{1}\\
		c_{n}^{2}
	\end{pmatrix} = 0.
\end{gather}
which simultaneously determines $\varepsilon_n(k_z)$. As a result, we find that the energy disperses with the moment parallel to the magnetic field according to	 
\begin{equation}
	\varepsilon_{n}(k_z)=\hbar v\cdot\mathrm{sgn}(n)\sqrt{\frac{\epsilon_{|n|}^{2}}{l_B^2}+k_{z}^{2}}, \quad
	n = \pm 0, \pm 1, \dots
	\label{eq:dispersion90}
\end{equation}
where $\pm0$ denotes the ground states for electrons and holes respectively with convention $\mathrm{sgn}(\pm 0)=\pm 1$.

We still have to solve the eigenproblem \eqref{eq:transverse_chi} to determine $\epsilon_n$. Let us decouple equations via
\begin{gather}
  \label{couple1}
      Q^\dag Q\chi_n^1=\left(\frac{\epsilon_n}{l_B}\right)^2\chi_n^1,\quad
   Q Q^\dag\chi_n^2=\left(\frac{\epsilon_n}{l_B}\right)^2\chi_n^2.
\end{gather}
Operators  $H^+=Q^\dag Q$ and $H^-=Q Q^\dag$ in~\eqref{couple1} form a hermitian couple.
Written explicitly
\begin{gather}
  \label{couple2}
  H^{\pm}=p_y^2+W(y)^2\mp2\partial_y W(y),
\end{gather}
each of them can be associated with the Hamiltonian of a system. For description of the systems with Hamiltonians given by~\eqref{couple2} one can introduce supersymmetry generators where $W(y)$ plays the role of superpotential. The problem can thus be reformulated in terms of supersymmetric quantum mechanics.

Theoretical grounds of supersymmetric quantum mechanics are well developed. In some cases they even allow to find the full exact spectrum of a system, using $Q^{\pm}$ in a way analogous to creation and annihilation operators are used in ordinary quantum mechanics, see Refs~\onlinecite{Gendenshtein1983} and \onlinecite{Gendenshtein1985}. Here we need the two important statements of the theory: i) if the supersymmetry is unbroken, the ground state of the system is non degenerate and is exactly zero and ii) the supersymmetry is broken when $W(+\infty)\cdot W(-\infty)>0$. This breakdown leads to the ground state with positive energy.

We see that our function $W(y)$, see Eq. (\ref{HQ}) gives rise to a system with broken supersymmetry and non-zero degenerate ground state.
To see this explicitly one can try to construct an exact ground state of the Hamiltonians $H^{\pm}$. E.g. for $H^{+}$ we have:
\begin{gather}
   Q\chi_0(y)=0\ \Leftrightarrow \partial_y\chi_0=W(y)\chi_0\Rightarrow\notag\\
   \chi_{0}(y) = \exp\left(\frac{\hbar}{6mvl_B}(y^{3} - 3k_{0}^{2}l_B^2y)\right).
\end{gather}
This solution although formally of zero energy,  is not normalizable (the zero mode of $H^-$ suffers from the same disease).

The spectrum of the problem can still be found analytically for $k_0l_B \gg 1$. In the limit of $k_0 \to +\infty$ with $\zeta$ fixed, the Weyl cones become separated and the $\psi$--function near each Weyl point is given\cite{Andreev2016} by an appropriate combination of Hermite functions $\psi_n^{\text{osc}}(y)$ with $\epsilon_n = \sqrt{2\zeta n}$. To solve the present problem, we introduce a coupling constant $g \equiv (4\zeta k_0^2l_B^2)^{-1}\ll 1$ and rescale $y \mapsto y/\sqrt{\zeta}$. Using Eq.~(\ref{couple1}) we write down the corresponding Schr\"{o}dinger equation
\begin{equation}
	\left[ -\partial_{y}^{2} + g \left(y^{2} - \frac{1}{4g}\right)^{2} \mp 2\sqrt{g}y \right] \chi^{1,2}
	=  2\nu\chi^{1,2},
	\label{eq:transverse_schrodinger_1}
\end{equation}
where we denoted $2\nu \equiv \epsilon_n^{2}/\zeta$. For $g \to 0$ we have two independent harmonic oscillators and eigenfunctions are approximately
\begin{gather}
   \begin{split}
    \chi_{n\neq0}^1(y) &\sim \psi_{n}^{\text{osc}}\left(y-\textstyle{\frac{1}{2\sqrt{g}}}\right) - (-1)^{\lfloor\frac{n}{2}\rfloor} \psi_{n-1}^{\text{osc}}\left(y+\textstyle{\frac{1}{2\sqrt{g}}}\right), \\
    \chi_{0}^1(y) &\sim \psi_{0}^{\text{osc}}\left(y-\textstyle{\frac{1}{2\sqrt{g}}}\right), \qquad
    \chi_n^2(y) = \chi_n^1(-y),
   \end{split}
\end{gather}
with eigenvalues $\nu_{n}=\lceil\frac{n}{2}\rceil$, $n \in \mathbb{N}_0$ (so that $\nu_0=0,\;\nu_1=1,\;\nu_2=1$ etc.)

Let us first discuss the zeroth level. Non-normalizable quasi-solution at zero energy is
\begin{equation}
\chi_{\epsilon=0}(y_+) = e^{ - \frac{y^2_+}{2} - \sqrt{g}\frac{y^3_+}{3} },\quad y_+ = y - \frac{1}{2\sqrt{g}}.
\end{equation}
From the one hand, it satisfies \eqref{eq:transverse_schrodinger_1} with $\epsilon = 0$. From the other hand, it can be expanded in powers of $\sqrt{g}$ to produce $n$--th order perturbation theory result for \eqref{eq:transverse_schrodinger_1} implying that perturbatively $\epsilon^{(n)}_0 = 0$.
In order to find correct non-perturbative energy levels, we resort to standard WKB technique (see Appendix for details) and obtain for the zeroth level
\begin{gather}  \label{levels}
   \epsilon_0 = \sqrt{\frac{\zeta}{\pi}}e^{-1/6g}.
\end{gather}
From here we recover the ground state energy~\eqref{ground}. Result~\eqref{ground} is valid up to $g < \frac{1}{4}$ (i.\:e. for magnetic field up to $B_0$) with error less then 3\%. Higher levels are shifted by anharmonicity of the potential and splitted according to
\begin{equation}
    \label{level1}
	\epsilon_{2k}-\epsilon_{2k-1} = \sqrt{\frac{\zeta}{\pi}}\left(\frac{2}{g}\right)^{k} \frac{e^{-1/6g}}{k!}.
\end{equation}
In Eq.~\eqref{levels} and~\eqref{level1}  only the leading term in semiclassical expansion is retained.

Finally, let us consider intermediate values of angle $\eta$. Since the spatial variables do not separate in this case, we computed the LL dependence on longitudinal (along magnetic field) momentum \(\varepsilon(k_{\parallel})\) numerically (see \figref{fig:landau_levels}). One observes an interesting crossover from a field-independent level at $\eta=0$, Eq. (\ref{eq:dispersion00}) to a level, weakly dependening on magnetic field, Eq. (\ref{eq:dispersion90}). The field dependence in the latter case is due to the gap between $\pm 0$ levels, resulting from supersymmetry breaking of the underlying Hamiltonian at $\eta=\pi/2$. Interestingly, this gap turns into the gap between zeroth and first LL as $\eta$ evolves from $\pi/2$ to $0$.

\section{Magnetoconductance}
We now evaluate the conductance in the presence of magnetic field perpendicular to \pn junction. We will make use of Landauer formalism and solve the scattering problem for electrons moving from conductance to valence band through \pn junction. As our discussion of the LLs suggests, the result will be qualitatively different for two orientations of the junction with respect to the nodes splitting: parallel $\eta = 0$ and perpendicular $\eta = \frac{\pi}{2}$. In both cases, the spatial variables can be separated for longitudinal and transversal motion in the magnetic field.

For $\eta=0$ in the transversal motion there exists a field-independent mode, see Eq. (\ref{eq:dispersion00}). After substitution $\psi\mapsto\frac{1}{\sqrt{2}}(1+i\sigma_{y})\psi$ and separation of variables $\psi^{1,2}(x,y)=\psi_n^{1,2}(y)\phi^{1,2}(x)$ with functions $\psi_n$ given by \eqref{eq:ll_parallel_psi}, the scattering problems reads
\begin{equation}
	\left[\frac{\hbar^2(\hat{k}_x^2-k_0^2)}{2m}\sigma_z + \frac{\hbar v}{l_B}\sqrt{2n}\sigma_x - |e|Ea\operatorname{th}\left(\frac{x}{a}\right)\right]\phi = \varepsilon \phi.
\end{equation}
Transmission coefficient for the zeroth level $n=0$ is field-independent and for $k_0a\gg1$ is $T_0\approx 1$ (slight suppression from unity is due to the scattering between the nodes induced by built-in electric field). After accounting for LL degeneracy this results in linear contribution to magnetoconductance, $G(B) \propto (e^2/h)(BS/\Phi_0)T_0$, $l_B \ll l_E$. Thus, for $\eta=0$ the presence of a second nearby Weyl node does not change magnetoconductance qualitatively, as long as built-in electric field does not transfer particles between the nodes.\\

The situation is very different for the junction perpendicular to $\mathbf{k}_0$. Longitudinal ($z$) and transverse ($x$,$y$) variables can be separated in Landau gauge $\bm{A}=(-By,0,0)$
\begin{equation*}
	\frac{\H}{\hbar v} = \frac{\zeta}{2k_0} \left(\left(k_{x} -	\frac{y}{l_{B}^{2}}\right)^{2} - k_{0}^{2}\right)\sigma_{x} + \sigma_{y}\hat{k}_y + \sigma_{z}\hat{k}_z + \frac{z}{l_{E}^{2}}.
\end{equation*}
Substitution $$\psi^{1,2} = e^{ik_xx}\chi^{1,2}(yl_B^{-1}-k_xl_B)\phi^{1,2}(zl_E^{-1})$$ leads to transverse equations
\begin{equation}
	\left[ \frac{\zeta}{2k_0l_B}\left(y^{2}-k_{0}^{2}l_B^2\right)\sigma_{x} -i\partial_{y}\sigma_{y} \right]
	\chi_n(y)	= \epsilon_n\chi_n(y),
	\label{eq:transverse}
\end{equation}
which is the same as \eqref{couple1}, and scattering problem
\begin{equation}
	\left[(-i\partial_{z})\sigma_{z} + (l_E/l_B)\epsilon_{n}\sigma_{x} + z\right]
	\phi(z)	= 0,
\end{equation}
which is equivalent the Landau-Zener one and the transmission coefficient $T = \exp\left(-\pi(l_E/l_B)^2\epsilon_n^2\right)$ is determined by energies $\epsilon_n$.

Producing summation over Landau levels
\begin{equation}
	G(B) = \frac{e^2}{h}\frac{S}{2\pi l_B^2} \sum\limits_{\epsilon_n} \exp\left[-\pi\frac{l_E^2}{l_B^2}\epsilon^2_n\right]
	\label{eq:sum}
\end{equation}
and taking into account Eq.~\eqref{ground}, we obtain dependence $G(B)$ depicted at \figref{fig:magnetoconductance}:
\begin{equation}
	G(B) = 2\pi G_0b\left[\exp\left(-be^{-\frac{4}{3}\frac{a}{b}}\right) + \frac{\exp(-\pi b)}{2\sinh(\pi b)}\right],
  \label{eq:cond}
\end{equation}
where we have introduced
\begin{equation}
	G_0 \equiv \frac{2}{\zeta}\frac{e^2}{h}\frac{S}{(2\pi l_E)^2},	\quad
	a \equiv (\zeta k_0 l_E)^2,	\quad
	b \equiv \zeta \frac{l_E^2}{l_B^2}.
\end{equation}
In the two--cone model magnetoconductance has a maximum at magnetic field
\begin{equation}
	\label{eq:B_crit}
  B_c \sim \frac{4\zeta}{3\pi}\frac{\Phi_0k_0^2}{\ln\frac{4}{3}(\zeta k_0 l_E)^2},
	\quad	\zeta k_0 l_E \gg 1.
\end{equation}

Eqs~\eqref{eq:cond} and \eqref{eq:B_crit}  are the main predictions of our paper. For a plot of magnetoconductance as a function of magnetic field \(G(B)\) see \figref{fig:magnetoconductance}. To test feasibility of the found results we take numerical values for TaAs \cite{Lv2015TaAs} and TaP \cite{Xu2016TaP,Zhang2017} and estimate critical parameters. We suppose doping is weak enough to estimate $E \simeq \Delta^2/(\hbar v)^{3/2}$ according to Ref.~\onlinecite{Andreev2016}.
\begin{table}[!h]
\centering
\caption{Numerical values, taken from Ref. \onlinecite{Lv2015TaAs} for TaAs, and Refs. \onlinecite{Xu2016TaP,Zhang2017} for TaP.}
\begin{tabularx}{.7\linewidth}{|>{\centering}X|c|c|}\hline
                            & \enskip TaAs (W2) \enskip	& \enskip TaP (W1) \enskip	\\\hline
	$2k_0$, \AA${}^{-1}$	&	0.0183	&	0.021	\\\hline
	$\Delta$, meV			&	2		&	$\lesssim2$		\\\hline
	$\zeta = v_x/v_y$ 		&	1.65	&	1.6		\\\hline
	$l_E$, \AA				&	470		&	720		\\\hline
	$\zeta k_0 l_E$			&	7		&	12		\\\hline		
	$B_0$, T				&	9		&	11		\\\hline
	$B_c$, T				&	3		&	3.7		\\\hline	
\end{tabularx}
\label{tab:numbers}
\end{table}

The numbers presented in \tabref{tab:numbers} show that the situation we consider is indeed possible in an experimental setup. As $B_c < B_0$, effective coupling constant $g$ corresponding to position of the maximum is indeed small and our WKB calculation is valid for such fields.
\section{Conclusions}
To conclude,  we have studied the LL structure in WSM analytically ($\eta=0,\pi/2$) and numerically ($0<\eta<\pi/2$). Our analytical results are summarized in Eqs. \eqref{ground} and \eqref{eq:crit} as well as \figref{fig:landau_levels}. We believe that the gap in the LL spectrum predicted by this equation at n = 0 has already been observed in experiment Ref.~\onlinecite{Zhang2017} (the authors explained it via numerical solution of the Schroedinger equation). Our study completes these findings via analytical solution and numerical description of the crossover from $\eta=0$ to $\eta=\pi/2$ with rotating magnetic field.

We have also described how the tunnelling between Weyl nodes leads to the change in the behavior of magnetoconductance of \pn junction in WSM. We found that this tunnelling leads to the appearance of the characteristic field $B_c$, Eq.~\eqref{eq:B_crit},
at which the differential magnetoconductance changes its sign. We believe the same feature would exist at intermediate angles, but due to the absence of separation of longitudinal and transversal motion at $0<\eta<\pi/2$ in the two-cone approximation we were not able to study this problem in more detail.

In our treatment, we have completely discarded the influence of disorder and interaction. It means that characteristic traversal time through a pn-junction should be smaller then the quasiparticle relaxation time. The transport relaxation time in TaAs was estimated in i.e. Ref.~\onlinecite{Zhang2016}, $\tau=7\cdot10^{-13}$ s and \(v_F\approx 0.5\cdot10^6\ m/s\). Then the width of the \pn junction should be less than $\sim 1\mu m$. We have also neglected the Zeeman splitting which is negligible as long as magnetic field is smaller than spin-orbit interaction scale which produces the spin-orbit splitting of the quasiparticle
bands. For TaAs the corresponding magnetic field is estimated\cite{Arnold2016Chiral,Ramshaw2017} to be around $50$ T. Therefore there exists plenty of space for purely orbital magnetic-induced tunnelling in the framework of low-energy Hamiltonian Eq. (\ref{eq:weyl_pair}).

Overall, we are positive that the undertaken analysis helps to shed some light on the structure of a realistic WSM in moderate and strong magnetic fields and hope that the predicted behavior of  magnetoconductance of \pn junctions is going to be measured in the coming experiments.


\appendix*
\section{Semiclassical computation of ground state energy of tilted supersymmetric double--well potential}\label{app:ss}
\setcounter{equation}{0}\renewcommand{\theequation}{A.\arabic{equation}}
Eigenlevels of the Schr\"{o}dinger equation
\begin{equation}
	\left[ -\frac{1}{2}\partial^{2} + \frac{g}{2} \left(y^{2} - \frac{1}{4g}\right)^{2} - \sqrt{g}y \right] \chi(y)
	=  \nu\chi(y)
	\label{eq:tilted_double_well}
\end{equation}
can be studied in the limit of $g \ll 1$ via semiclassical approximation. Proper quantization condition taking into account both perturbative and non-perturbative corrections in small $g$ can be derived via uniform WKB\cite{Dunne2014}. Non-perturbative corrections to zeroth energy level (where they are the only ones), as well as to higher levels (fully determining the LL level splitting) can be found in the following way.

Near the minimum $y_\pm = y \mp \frac{1}{2\sqrt{g}}$ potential is quadratic ($\nu_+ \equiv \nu$, $\nu_- \equiv \nu - 1$)
\begin{equation*}
	\left[ -\partial^{2} +  y_\pm^2 - (2\nu_\pm + 1) \right] \chi(y_\pm) \approx 0,	\quad	|y_\pm| \ll \frac{1}{\sqrt{g}}.
	\label{eq:approx_hermite}
\end{equation*}
Decaying at $y \to +\infty$ solution is given by Hermite function $\psi_\nu^{\text{osc}}(y)$ which asymptotes
\begin{equation*}	
	\psi_\nu^{\text{osc}} \sim
	\begin{cases}
		(2y)^\nu e^{-\frac{y^2}{2}},	&	y \to +\infty,	\\
		\cos{\pi\nu}(-2y)^\nu e^{-\frac{y^2}{2}} + \dfrac{\sqrt{\pi}e^{\frac{y^2}{2}}}{\Gamma(-\nu)(-y)^{\nu+1}},	&	y \to -\infty.	 	\\
	\end{cases}
	\label{eq:parabolic_asymptotics}
\end{equation*}

We have to match $\psi_\nu^{\text{osc}}$ with WKB solution, valid under the potential hump $|y| \sim \frac{\const}{\sqrt{g}}$
\begin{equation}
	\chi = C_+\frac{\exp\int_{0}^y |k(z)|dz}{\sqrt{|k(y)|}} + C_-\frac{\exp-\int_{0}^y |k(z)|dz}{\sqrt{|k(y)|}},
	\label{eq:semiclassics}
\end{equation}
\begin{equation}
	|k(y)|^2 = g\left(y^2 - \frac{1}{4g}\right)^2 -2\sqrt{g}y  - 2\nu.
\end{equation}
To this end, we expand semiclassical action at $|y| \lesssim \frac{1}{2\sqrt{g}}$
\begin{multline*}
	S(y) = \frac{1}{8g}\int_{0}^{2\sqrt{g}y}\sqrt{(1-z^2)^2 -16g(z+2\nu)}\,dz \sim \\
		 \sim \left.\left[\frac{1}{8g}\left(z-\frac{z^3}{3}\right)
		  + \frac{1}{2}\ln\left|1-z^2\right| + \nu\ln\left|\frac{1-z}{1+z}\right|\right]\right|_{0}^{2\sqrt{g}y}
\end{multline*}
and near the minimum $|y_\pm| \ll \frac{1}{\sqrt{g}}$
\begin{align*}
	S(y) &\sim \frac{1}{12g} - \frac{y_+^2}{2} + \frac{1}{2}\ln\left[-4\sqrt{g}y_+\right] + \nu\ln\left[-\sqrt{g}y_+\right],\\
		 &\sim -\frac{1}{12g} + \frac{y_-^2}{2} + \frac{1}{2}\ln\left[4\sqrt{g}y_+\right] - \nu\ln\left[\sqrt{g}y_-\right]
\end{align*}
where we neglected terms $\bigo{y_\pm^3}$. Matching \eqref{eq:semiclassics} with asymptotics of Hermite functions we derive the quantization condition
\begin{equation}
\left(\frac{g}{2}\right)^{2\nu} \Gamma(\nu)\Gamma(1+\nu)\tan^2\pi\nu = \frac{\pi}{2}e^{-\frac{1}{3g}},
\end{equation}
which for $\nu \to 0$ becomes $\nu_0 = e^{-\frac{1}{3g}}/2\pi$ and for $\nu \to \mathbb{N}_+$ gives energy level splittings due to inter-well tunnelling.
 These results have been derived via instanton technique in Refs. \onlinecite{Balitsky1986,Zinn-Justin2004}.

\begin{figure}[!h]
	\centering
	\includegraphics[width=.9\columnwidth]{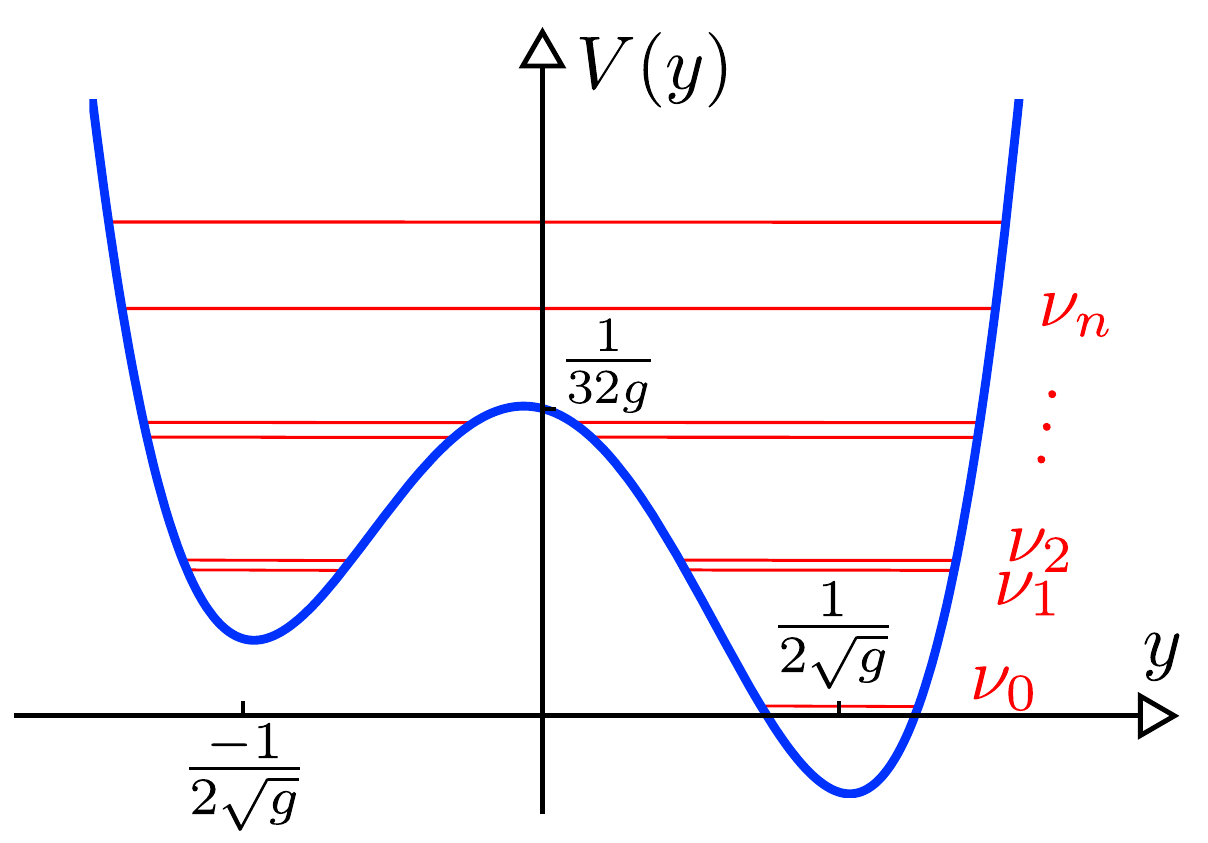}
	\caption{Tilted supersymmetric double--well potential \eqref{eq:tilted_double_well} and its energy levels $\nu_n$ with coupling constant $g=2^{-6}$.}
	\label{fig:tilted_double_well}
\end{figure}

\bibliography{weyl}

\end{document}